\documentclass[12pt,preprint]{aastex}

\def\gapprox{\;\rlap{\lower 3.0pt                       
             \hbox{$\sim$}}\raise 2.5pt\hbox{$>$}\;}
\def\lapprox{\;\rlap{\lower 3.0pt                       
             \hbox{$\sim$}}\raise 2.5pt\hbox{$<$}\;}

\def\deut{$^{2}$H }
\def\lisix{$^{6}$Li }
\def\lisev{$^{7}$Li }
\def\benine{$^{9}$Be }
\def\nlia{$N($Li$)$ }
\def\nfe{$N($Fe$)$ }
\def\nli7{$N(^{7}$Li$)$ }
 
\begin{document}
 
\title{A Critical Examination of Li Pollution and Giant Planet
Consumption by a Host Star}

\author{Eric L. Sandquist, Jon J. Dokter} 
\affil{San Diego State University, Department of Astronomy, San Diego,
CA 92182} 
\email{erics, dokter@mintaka.sdsu.edu}

\author{D.N.C. Lin}
\affil{Department of Astronomy \& Astrophysics, University of California, 
Santa Cruz, CA 95064}
\email{lin@ucolick.org}

\author{Rosemary A. Mardling}
\affil{School of Mathematical Sciences,
Monash University, Melbourne 3800, Australia}
\email{mardling@monash.edu.au}

\begin{abstract}

Given the high metal contents observed for many stars with planets
(SWPs), we examine the overall likelihood that the consumption of a
giant planet could pollute its host star. First, we discuss \lisix and
\lisev as indicators of pollution, verifying that \lisix is a strong
indicator of pollution 30 Myr after star formation, and showing that
it strongly constrains the amount of heavy element pollution
incorporated into the star. Detection of \lisix in SWPs could also be
used to distinguish between giant planet formation theories, and can
be used to detect the consumption of giant planets independent of
planet mass. Second, we examine the probability that giant planets
between 1 and 3 $M_{J}$ could be destroyed in the outer convection
zone of stars slightly more massive than the Sun (for which detection
of a chemical signature of pollution would be easiest). We find that
heated giant planets would be efficiently destroyed near the surface
of the star, while the cores of cold giant planets may be able to
survive a plunge through the base of the star's convection
zone. Third, we examine whether dynamical processes could bring a
giant planet close enough to the star to destroy it, and whether the
destruction of a planet would necessarily affect other planets in the
system. While tidal interaction between protoplanets and their nascent
disks may have led them to the proximity of their host stars,
post-formation star-planet interaction can lead to tidal disruption of
the planet and accretion of its material, or orbital decay followed by
hydrodynamical interaction.  Throughout, we consider the case of HD
82943, a star known to have two planets and having a preliminary
detection of \lisix.  Using stellar models including diffusion, we
estimate the mass of the HD 82943 to be $\sim 1.2 M_{\odot}$ and its
age to be $\sim 0.5-1.5$ Gyr. The observed \lisev abundance for HD
82943 is consistent with stars of similar $T_{eff}$ and age in the
open cluster NGC 752.  We describe a possible dynamical history for a
hypothetical planet in the presence of the two resonant planets
currently known.  We present stable orbital configurations in which
the hypothetical planet has low eccentricity and semi-major axis near
0.02 AU, so that it is dynamically decoupled from the resonant
planets.  Tidal interactions with the slowly-rotating star can
subsequently drag the planet into the stellar surface within the age
of the star.
\end{abstract}


\section{Introduction}
The discovery of extrasolar planets in the past few years (starting
with Mayor \& Queloz 1995) has led to the realization that they are
common around solar-type main sequence stars (for a review, see Marcy
\& Butler 1998). Most of the discovered planets have masses comparable
to that of Jupiter and Saturn, but have orbital characteristics (such
as short period or large eccentricity) that are very different from
those of the giant planets in the solar system.  Some of these
properties result from the circumstances of their formation, while
others are determined by post-formation dynamical evolution.  An
important challenge is to isolate the dominant cause behind the
orbital properties of the population of planetary systems.
 
Nearly all extrasolar planets have been found with radial velocity
surveys of nearby stars, and in order to optimize the detection
efficiency, these surveys have been focused on mature solar-type stars
with quiet atmospheres.  A particularly interesting characteristic of
the planet-bearing stars is that they tend to be metal rich with
respect to the Sun and the F-G field star average in the solar
neighborhood (e.g. Gonzales et al. 2001; Santos, Israelian \& Mayor
2001).  Two explanations have been advanced to explain the association
of metal-rich stars with planets.  First, metal-rich planetesimals or
planets could have been consumed in the outer convective layers of the
host star during the process of planet formation (e.g. Gonzalez 1997;
Ford, Rasio, \& Sills 1999). Our main focus for this paper is
planetary consumption following migration through the protoplanetary
disk (Lin 1997; Laughlin \& Adams 1997; Sandquist et al. 1998) or
gravitational interaction with other planets (Rasio \& Ford 1996).
Even though the contaminants are expected to be thoroughly mixed
within the convective envelope, this surface layer is sufficiently
shallow for F-G stars that the consumption of one or a few
Jupiter-like planets would be adequate to account for observed
metallicity enhancements above the field star average. Second,
enhanced metallicity in the planet-forming disk (probably resulting
from a metal-rich parent cloud) might be especially conducive to
planet formation.  In fact, Pollack et al. (1996) show that the timescale for giant planet formation in the disk (by the conventional
accretion of gas onto a solid core) decreases rapidly with the surface
density of solid material, which presumably derives from metal content
of the gas.  Thus, the first scenario attributes the excess metal
content of planet-bearing stars to the dynamical evolution of the
planet, whereas the second scenario assumes it to be associated with
planet formation.

Detailed analysis of the chemical composition of stars with planets
may hold the key to determining the relative importance of these two
possibilities (Santos, Israelian, \& Mayor 2000, Gonzalez et
al. 2001). For example, light element abundances can constrain the
time of pollution.  This conjecture has recently been spotlighted by
the apparent discovery of \lisix in the spectrum of the star with
planet (SWP) HD 82943 (Israelian et al. 2001).  Together with Be and
B, \lisix is produced primarily through cosmic ray nucleosynthesis.
Nuclear burning destroys \lisev at $\sim 2.5 \times 10^6$ K and \lisix
at even lower temperature. Because HD 82943 is a mature main sequence
star, the temperature at the base of its surface convection zone is
low enough to preserve any remaining \lisix and \lisev.  However,
during the pre-main sequence phase of its evolution, the temperature
at the base of the more extended convection zone is large enough to
completely destroy surface \lisix, and deplete \lisev.  In \S2, we
briefly discuss our own stellar models of SWPs, and examine the
possibility of detecting any signature of a pollution mechanism that
may operate while a star is on the main sequence by using observations
of \lisix and \lisev. We also discuss the parameters of the SWP HD
82943.

As described above, one possible avenue for the late accretion of Li
isotopes is the consumption of a giant planet.  In \S3 we present new
numerical simulations to examine the possibility of dissolving a giant
planet in the convective envelope of a SWP.  This calculation is an
extension of our previous calculations to the limiting case of F stars
that have relatively low-mass convection zones.  Our primary interest
is to determine whether a Jupiter-mass planet can avoid total
disintegration after entering into the atmosphere of a host star of
that type.  In \S4, we consider the dynamical origin of such a
planet. We find one possible scenario for bringing the giant planet
into contact with its host star that involves disk migration of the
giant planet to an orbital period slightly less than the minimum
observed for known extrasolar planets. Subsequent internal tidal
dissipation causes this hypothetical planet to dissipate its orbital
energy.  If the planet reaches the proximity of the host star with a modest
eccentricity, tidal dissipation may cause it to overflow its Roche
lobe and become tidally disrupted.  If the planet's initial orbit is
essentially circular, it would undergo moderate tidal interaction with
its host star.  We show that it is possible for a planet with an
initial period comparable to that of known short-period planets to
spiral into the envelope of a SWP within 1 Gyr.  We also use HD 82943
as an example to demonstrate the existence of initial dynamically
decoupled configurations in which the observed kinematic properties
of known planets may be preserved despite the presence and demise of
doomed short-period planets.  Finally in \S5, we summarize our results
and discuss their implications.

\section{Stellar Models}

To produce stellar structure models for input into our hydrodynamics
code (see \S3), we have used the one-dimensional stellar evolution
code developed by Eggleton (1971, 1972), with input physics updated by
Pols et al. (1995). Our implementation of the code includes recent
opacities for the surface (Alexander \& Ferguson 1994) and interior
(Iglesias \& Rogers 1996), and reaction rates (Bahcall \& Pinsonneault
1995). Chemical diffusion has also been incorporated using rates
calculated using the method of Thoul, Bahcall, \& Loeb (1994). Our
intent is to ensure that we are modeling the stellar structure and
convection zone depth as accurately as possible with current
physics. We have tested the accuracy of our numerical methods by
obtaining the sound speed and density residuals of our standard solar
model using the radial profiles derived from helioseismic data by
Basu, Pinsonneault, \& Bahcall (2000). As found by Basu et al., models
that do not include diffusion are much poorer matches to the Sun's
density and sound speed profiles, particularly near the base of the
convection zone. Typical residuals for our models that included
diffusion were about 0.4\% in sound speed and 3\% for density.  We
have also compared the radius of the convection zone boundary and
surface helium abundance of our standard model with the values of the
diffusion-inclusive solar model of Brun, Turck-Chi\`{e}ze, \& Morel
(1998), and again found satisfactory agreement.

\subsection{Light Element Abundances in Solar Type Stars}\label{li6}

Our stellar evolution code has also been modified to follow abundances
of light elements \lisix, \lisev, and \benine.  Before examining other
details of the pollution process, we will discuss these chemical
tracers and their observability in SWPs.  As many others have noted
previously, the absolute value of \nlia can be used to distinguish
substellar objects from stars (see Chabrier \& Baraffe 2000 and Basri
2000 for recent reviews), and because Li is preserved in these
objects, stars could well be enriched by pollution mechanisms
(Alexander 1967).  However, evidence from open clusters indicates that
it is apparently depleted by single-star mechanisms on timescales of
$\sim 300-600$ Myr for main sequence stars with effective temperatures
at or slightly above that of the Sun (see Figs. 3 and 4 of
Pinsonneault 1997). Thus, the use of \nlia to look for signs of
pollution requires a fairly accurate estimation of the age of a
SWP. Scatter in measured \nlia values in open clusters makes
definitive detection of a pollution signal even more difficult.

Because \lisix burns at a lower temperature than \lisev, models show
that primordial \lisix is destroyed while the star is still on the
pre-main sequence (before an age of about $3 \times 10^{6}$ yr;
Forestini 1994). However, for stars with masses comparable to that of
the Sun, the subsequent retreat of the convection zone base means that
surface material is no longer exposed to temperatures high enough to
deplete \lisix. This evolutionary process places \lisix in a unique
position: species like \deut which burn at lower temperatures would be
depleted rapidly even on the main sequence, while \lisev and \benine
are not depleted by large amounts on the pre-main sequence, so that
initial abundance variations and not-yet-understood mixing processes
complicate the interpretation of abundances. \lisix is the equivalent
of an on-off switch for a range of star masses near $1 M_{\odot}$.
(Because convection-zone mass increases steeply with decreasing
stellar mass, the dilution of pollutants is an important consideration
for stars less massive than the Sun, and so detection depends on the
sensitivity of observational techniques. For stars of solar
metallicity below $1 M_{\odot}$, the mass in the convection zone can
be approximated as $M_{cz} / M_{\odot} \simeq 0.11 (1 - M / M_{\odot})
+ 0.019$. For stars more massive than the Sun, detectability of
pollution depends on details of the pollution process: for example,
what fraction of the pollutants penetrate below the convection zone of
the star. We will discuss this issue in the last paragraph of this
subsection.) If detected in significant amounts in a SWP, \lisix would
provide definitive evidence of a pollution process in effect after an
age of about $10^{7}$ yr.

Because \lisix is not depleted after $10^{7}$ yr, the amount of \lisix
detected can be used to constrain the amount of polluting material
(Israelian et al. 2001). However, it should be noted that the amount
of {\it heavy element pollution} can probably be much more accurately
constrained than total pollution. The abundance of \lisix and \lisev
in solar system meteorites is measured relative to other heavy
elements that have well-determined abundances relative to hydrogen in
the Sun.  After correcting these comparison heavy elements to a solar
composition, the abundance is determined to be $\log N_{Li} = 3.31 \pm
0.04$ ($\log N_{H} = 12.00$; Anders \& Grevesse 1989), while that of
\lisix is found to be an order of magnitude smaller. The more directly
observed \lisev abundance in the interstellar medium yields a similar
value (Lemoine et al. 1993; Knauth et al. 2000).  Using the assumption
that the Li/H ratio in the interior of giant planets is the same as
the corrected meteoritic value, Israelian et al. (2001) infer that for
SWP HD 82943 a giant planet of mass $1-3 M_{J}$ could have provided
the \lisix necessary to pollute the shallow convective envelope of the
star.

However, this inference is probably an overestimate of the amount of
giant planet material needed to account for the \lisix measurement. As
indicated above, the nature of meteoritic abundance measurements makes
the relative abundance of Li and Fe more accurate than the relative
abundance of Li and H.  Independent of that, many planetesimals or one
giant planet could pollute a star by the same amount --- the measured
\lisix places no direct constraint on the hydrogen (and helium)
content of the polluter.  As a result, it is more accurate to use a
measured \lisix abundance to calculate the amount of heavy elements
consumed.

The mechanism by which giant planets form can have an important
bearing on \lisix pollution though. If planets are formed by
gravitational instabilities in the protoplanetary disk (hereafter GI;
see Boss 2001 for a recent reference), then there should be a roughly
linear relationship between \lisix pollution amount and consumed giant
planet mass (and the solar Li/H ratio is appropriate to use in
computing total pollutant mass). If giant planets are instead
nucleated around a rock/ice core which accretes gas after it reaches a
critical mass of $10 M_{\oplus}$ or more (hereafter CA; Podolak,
Hubbard, \& Pollack 1993), the \lisix pollution mass would also be a
linearly increasing function of the planet mass, but with non-zero
pollution for low giant planet masses. The necessity of a solid core
in this case implies that the average metal content for a giant planet
will be higher than that of the gas used to form the star. Even in the
absence of a core structure today, high metal content is still needed
to account the relatively compact sizes of solar system giant planets.
Models of Jupiter indicate that $0.03 < \overline{Z} < 0.14$ (Guillot
et al. 1997).

For a giant-planet core of $15 M_{\oplus}$, one would expect
approximately $4 \times 10^{44}$ \lisix atoms, assuming a constant
\nlia/\nfe ratio ($6.3 \pm 0.6 \times 10^{-5}$), $X_{Fe} / Z = 0.07$,
and \lisix / \lisev = 0.08 (all from meteoritic measurements; Anders
\& Grevesse 1989). The importance of this mass estimate is that a
giant planet of nearly {\it any} mass could provide the necessary
lithium. Planets of larger mass would contain significantly more
lithium because accreted gas still has cosmic abundances. However,
because of the overwhelming abundance of hydrogen and helium in
accreted gas, that gas would generally be more than an order of
magnitude less rich in heavy elements. As a result, \lisix can provide
a means of looking for consumed planets that is nearly independent of
planet mass.  If indeed the planet mass function rises steeply toward
low masses like $M^{-1}$ (Marcy \& Butler 2000; Jorissen, Mayor, \&
Udry 2001), then the traces of destroyed low-mass giant planets are
the ones most likely to be found.  Low-mass planets are also more
likely to be scattered by gravitational interactions with other
planets within a given system.  If the mass of the host star's
convection zone can be estimated via stellar models, this inference
could provide a telling signature of giant planet consumption.

It is reasonable to ask whether post-formation accretion of
planetesimals or planets could return the \lisix abundance to near
primordial levels without noticeably over-enriching the star in
\lisev.  The best measurements of the \lisev / \lisix ratio come from
solar system meteorites ($\sim 12.5$; Anders \& Grevesse 1989) and
from clouds in the interstellar medium. Interstellar clouds appear to
show significant variations in \lisev / \lisix, ranging from near
solar values toward $\rho$ Oph, $\zeta$ Oph, and $\zeta$ Per
($11.1\pm2.0$, $8.6\pm0.8$, and $10.6\pm2.9$, respectively), to low
values seen in another cloud toward $\zeta$ Oph ($\sim 1.4$) and two
toward o Per ($3.6\pm0.6$ and $1.7\pm0.3$) [see Lemoine, Ferlet, \&
Vidal-Madjar (1995) and Knauth et al. (2000)]. If the low ratio values
seen toward the interstellar clouds are preserved in planetesimals,
effective enrichment in \lisix is possible without substantially
increasing the \lisev abundance.

The discussion so far has assumed that the entire planet is destroyed
either immediately outside the host star (and then accreted) or in its
convection zone --- if this is not the case, the uncertainty increases
again because the initial abundance and present distribution of Li and
other metals within gaseous giant planets are poorly constrained by
observations. The abundance of both \lisev and \lisix in the upper
atmosphere of Jupiter is too small to be detected because it is
depleted via the formation of the molecule Li$_{2}$S, which condenses
under pressure at $\sim 10^3$K (Fegley \& Lodders 1994, Lodders 1999).
The distribution of Li within a giant planet can play a role in some
pollution scenarios because the planet's envelope can shield the core
from dissolution until late in its plunge into a star. If Li is
uniformly distributed in the planet, a substantial amount will be
deposited into the star's convection zone. However, if Li is held
primarily in the core, it could be almost entirely deposited in the
radiative interior where it would not be observable. At this time, it
cannot be ruled out that a large fraction of a giant planet's Li is
contained in the core or deep atmosphere of the planet (Fegley,
private communication). According to the CA theory, these portions of
the planet form (along with meteorites) through the coagulation of
solid condensates that hold most of the lithium in the protostellar
disk. A major question is whether core material is re-melted and
redistributed throughout the planet's envelope by convection. As a
result of these uncertainties, it is important to examine {\it how} a
giant planet is destroyed during its collision with the host star. We
attend to this issue in \S3.

\subsection{The Case of HD 82943}

Because of the importance of the claimed detection of \lisix in HD
82943, we devote the discussion below to an examination of the
characteristics of the star. Because the \lisix detection depends on a
delicate fit of synthetic spectra to the shape of a much stronger
\lisev line and because the Israelian et al. (2001) measurements have
not yet been confirmed by an independent analysis, we must leave open
the possibility that the measurement might be disputed.  Nevertheless,
it is useful to consider the theoretical implications of any SWP with
a relatively high \lisix abundance, including HD 82943.

\subsubsection{Age and Metallicity}

The observations we used to constrain our models of HD 82943 are
summarized in Table~\ref{obs}. We list surface gravity as a potential
constraint, although we elected not to use it. First, an error bar of
$\pm 0.10$ results in a large uncertainty ($\sim 25$\%) in $M / R^{2}$
(Ford, Rasio, \& Sills 1999). Second, there is some indication that
there are significant systematic differences in values measured by
different groups (Gonzalez et al. 2001).

The relatively high effective temperature and metallicity of HD 82943
required us to consider masses near $1.20 M_{\odot}$, which is at the
high end of the range quoted by Israelian et al. More recent analysis
by Santos et al. (2001) supports this idea (they quote $M = 1.15 \pm
0.05 M_{\odot}$). We have checked this estimate by testing the output
of our code against observed values for the SWP HD 209458, which has
very similar observational characteristics.  Using a mass of $1.12
M_{\odot}$ derived by Gonzalez et al. (2001) from metal-rich models of
Schaerer et al. (1993), we are able to match $M_{V}$, $T_{eff}$, and
$\log g$ to within the quoted errors at an age of 2.5 Gyr (also within
the errors of their quoted age estimate).

\begin{figure}
\plotone{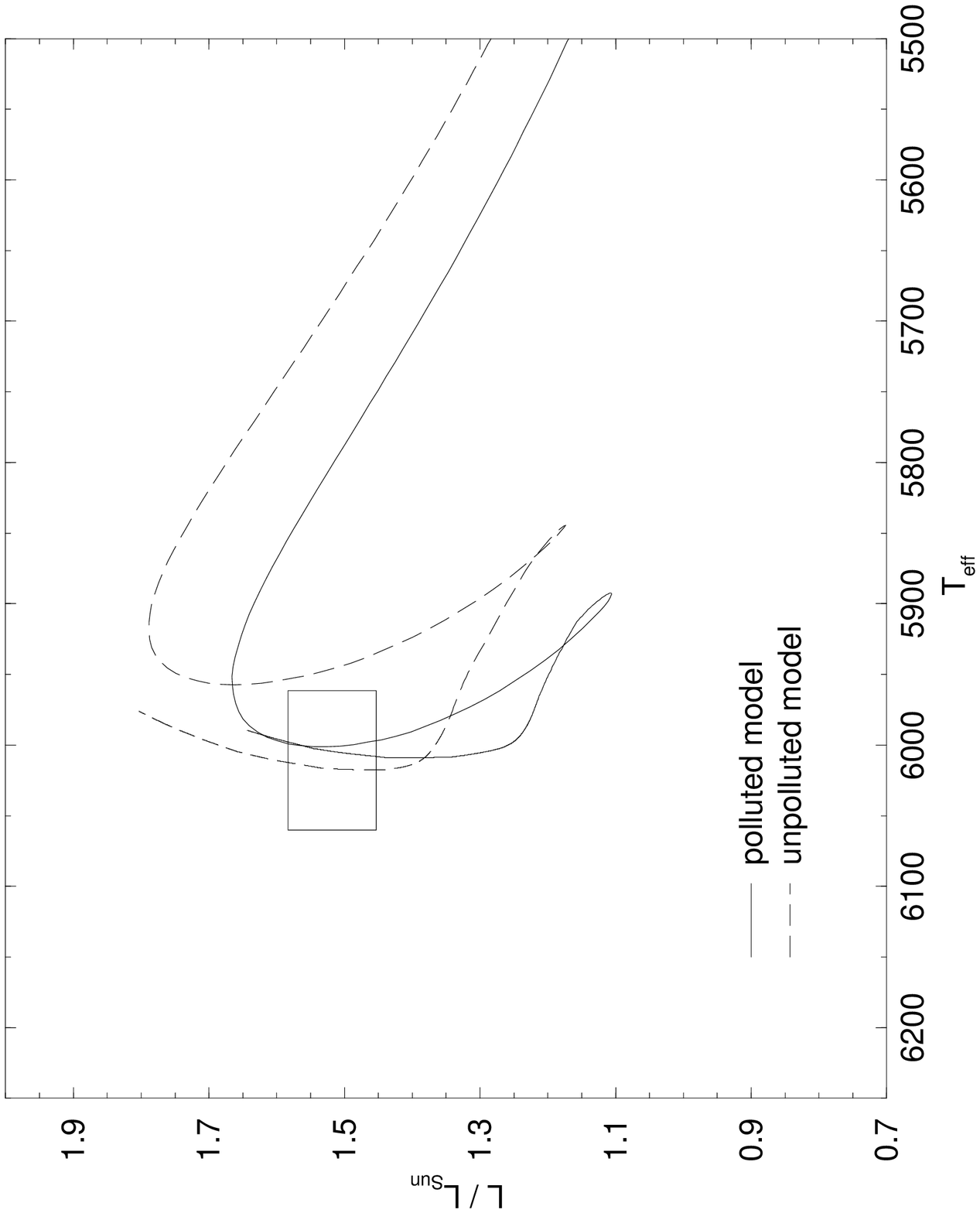}
\caption{The HR Diagram for our evolutionary models of HD 82943. The
box indicates the 1 $\sigma$ error bars for HD 82943 based on the data
in Table~\ref{obs}. \label{hrtrack}}
\end{figure}

Details of the two models for HD 82943 that we will discuss are
presented in Table~\ref{smods}. (Detailed estimation of the errors in
the mass and initial metal abundance are beyond the scope of this
paper.) Figure~\ref{hrtrack} shows the evolutionary tracks for our
models for comparison with the error box for HD 82943. The
``Unpolluted Model'' is a standard model including diffusion with a
uniform initial [Fe/H] throughout the interior. The surface convection
zone for this model contains a mass approximately half the mass of the
Sun's convection zone.

The observed metallicity for HD 82943 is at the high end of the the
field star distribution for G and early K field dwarf stars (Favata et
al. 1997). For this reason, we considered the possibility that the
outer layer of HD 82943 was contaminated by heavy elements from the
accretion of planetesimals or terrestrial planets after its formation.
Pollution would raise [Fe/H] in the convection zone, but would leave
the deep interior with a metallicity lower than the convective surface
layers (although diffusion will transport some of the convection
zone's metals into the interior).  Assuming pollution by $60
M_{\oplus}$ of heavy elements starting around the time the star
reached the end of the Hayashi track and continuing at a constant rate
for $10^{8}$ yr, we obtained a good candidate model with $M = 1.14
M_{\odot}$ (labelled ``Polluted Model'' in Table~\ref{smods}).  This
post-formation addition would result in increased lithium abundances
that can be ruled out by the observations. Earlier pollution would
probably have been diluted to a degree that it would not have been
observable. In addition, $60 M_{\oplus}$ of heavy element pollution
can probably be ruled out based on an examination of higher-mass SWPs
(Pinsonneault, DePoy, \& Coffee 2001; Murray \& Chaboyer 2002) and
limits on metallicity variation among solar-type stars in the Pleiades
cluster that have a dispersion in lithium abundance (Wilden et
al. 2002). Nevertheless, we conducted the calculation as a test of the
extremes possible in the envelope of HD 82943. With [Fe/H]$_{f} =
+0.22$, the consumption of a giant planet could boost the metallicity
of the convection zone (CZ) to near the observed value of $+0.32$.

We note that our models indicate that an age of 6 Gyr for HD 82943 (as
quoted by Israelian et al.)  is very unlikely. Our unpolluted model
falls within the HD 82943 error box between an age of 0.5 and 1.3 Gyr,
while the polluted run falls in the box between ages of 1.4 and 2.2
Gyr. The isochrones of Schaerer et al. (1993) also put the age of HD
82943 between about 1 and 2 Gyr. In addition, both of our models
require HD 82943 to have started with a super-solar metallicity. So,
independent of our assumptions about pollution after the pre-main
sequence, this seems to indicate that HD 82943 had to have a high
initial metal abundance, supporting the scenario that high metallicity
in star-forming gas may enhance the probability of planet formation
(see \S1).

\subsubsection{Lithium Abundance}

Israelian et al. (2001) compared HD 82943 with the open cluster M67
(age $\sim 4$ Gyr), and found that HD 82943's \lisev abundance was at
the high end of the distribution of stars from that cluster,
consistent with pollution.  However, the error in their quoted age
makes such an analysis misleading. Since \lisev is continuously being
depleted on the main sequence, a more realistic comparison would be
with stars from the younger cluster NGC 752 (Hobbs \& Pilachowski
1986) which has an age ($\sim 1.7$ Gyr) similar to that of HD 82943.
The abundance of HD 82943 is consistent with the mean trend in Li
abundances for NGC 752.  Although this similarity weakens the case for
\lisev enrichment (see also Ryan 2000), it does not rule out the
possibility that HD 82943 has been enriched in \lisev.

In \S~\ref{li6}, we have already argued that more constraints may be
placed on pollution by \lisix measurements. Using the estimate of $3.2
\times 10^{44}$ \lisix nuclei in the convection zone of HD 82943
(Israelian et al. 2001), we find that the corresponding mass of heavy
element pollutants should be near $13 M_{\oplus}$, consistent with
theoretical values for the critical core mass of giant planets
(Podolak et al. 1993).  For the assumptions, only about $1 M_{\oplus}$
of iron would have been consumed, which would not have noticeably
changed the star's heavy-element abundance ($\sim 0.03$ dex).

\section{Hydrodynamical Model of Star-Planet Interaction}

We next consider the process of contaminating the stellar convective
envelope via planet consumption.  In the next section, we suggest that
ultra-short-period planets may migrate toward their host stars as a
consequence of the tidal interaction between them.  Such a planet may
either disintegrate before it reaches the surface of the host star or
it may spiral into the star's atmosphere and envelope.  For the first
case, the planet's tidal debris may form an accretion disk or a swarm
of much smaller bodies.  Through mutual collisions and the secular
perturbations of other planets, the debris fragments would fall onto
the surface and become dissolved into the convective zone of the host.

For the second possibility, below we discuss the details of the
hydrodynamical interaction between a short-period giant planet and the
envelope of the host star. Here we model the degree to which the
planet can be ablated and dissolved before it passes through the base
of the convection zone of the star. For planetary disruption in the
deeper radiative zone, the deposited pollutants become essentially
unobservable.

\subsection{Numerical Scheme and Initial Conditions}

The hydrodynamics code used to do the calculations is largely
identical to that used in Sandquist et al. (1998). Briefly, the
computational domain was three-dimensional and composed of a main grid
and a nested subgrid, both of $64 \times 64 \times 64$ zones.  The
size of the subgrid was chosen to be comparable to the size of the gas
giant planet, and the main grid was then taken to be approximately 4
times larger in each dimension. Each simulation follows the
center-of-mass frame of the planet, which was initially taken to be on
a circular orbit at the star's surface. The gas that flows onto the
grid was given densities taken from the stellar evolution models
described earlier. Aerodynamical drag causes the planet to move
relative to the grid, and the adjustments the code makes to keep the
planet at the center of the subgrid are used to update the radial
distance from the center of the star, and the radial and tangential
velocities.

In our previous simulations (Sandquist et al. 1998), we examined the
interactions between cold Jupiter- and Saturn-mass giant planets and
stars of 1 and 1.22 $M_{\odot}$. The simulations presented here were
carried out to assess the feasibility of polluting HD 82943 with a
giant planet. However, these simulations have broader interest because
stars near this mass have easily polluted low-mass convection zones
with a large enough spatial extent that destruction of the planet in
the convection zone is still likely. In addition, we have examined the
effects that larger planet mass and ``inflated'' planet structure have
on the survivability of the planet. The discovery of an unexpectedly
large-size planet transiting HD 209458 (Henry et al. 2000, Charbonneau
et al. 2000) has made the last issue especially relevant.

In our simulations, we chose several planet models, which included: a
cold compact Jupiter-mass planet with a size $R_{p} = R_{J}$, a hot
inflated Jupiter-mass planet with a size $R_{p} = 1.4 R_{J}$ similar
to the planet orbiting HD 209458 ($R_{p} = 1.347 \pm 0.060 R_{J}$,
Brown et al. 2001; $1.321 \pm 0.007 R_{J}$, Hubbard et al. 2001), a
cold compact ($R_{p} = 1.12 R_{J}$) $3 M_{J}$ planet, and a hot
inflated $3 M_{J}$ planet ($R_{p} = 1.4 R_{J}$).  There are several
physical effects that may lead to the inflation of a short-period
planet.  For example, models of giant planets that develop near their
host stars which include the effects of stellar insulation predict
that the planets gravitationally contract less rapidly than they would
if they had evolved much farther away (Burrows et al. 2000).  During
the course of spin synchronization, orbital circularization, and
decay, tidal heating of short-period planets is also expected to
maintain a larger planet radius, or inflate a planet that started out
as cold and compact (Bodenheimer, Lin, \& Mardling 2001). A larger
radius for a planet results in a larger cross section, increased
aerodynamical drag, and quicker orbital evolution, but also lower
envelope binding energy and ease of ablation by the stellar gas.

In our hydrodynamic simulations, we have started the planet at the
surface of the host star in order to model the last stages of the
star-planet interaction. (The dynamical origin of this configuration
will be discussed in the next section.)  All four planets were modeled
as $n=1$ polytropes (Hubbard 1984).  The core structure of giant
planets is still poorly understood (particularly whether rock/ice
cores are present, and whether the existence of a core is affected by
heating mechanisms), although this issue has a relatively minor effect
on the overall structure of the planet (see Fig. 2 of Bodenheimer et
al. 2001). The structure of the $3 M_{J}$ planet is worth more
discussion, given that models are not constrained by observations as
yet. The overall planetary radius is the most uncertain quantity,
primarily because of uncertainties in albedo (e.g., Marley et
al. 1999). The indications from more detailed models are that the
structures (most importantly, the dependence of density on radius) of
planets with different total radii are likely to obey a rough homology
relationship. For the cold $3 M_{J}$ planet, we estimated the radius
using the models of Saumon et al. (1996). Based on the insensitivity
of giant planet radii to mass near $1 M_{J}$ (see Figs. 4 and 6 of
Saumon et al.) and the desire to make useful comparisons with our
Jupiter-mass runs, we chose to give our ``hot'' $3 M_{J}$ model the
same radius as our ``hot Jupiter'' model ($1.4 R_{J}$). This degree of
inflation for a $3 M_{J}$ planet would require a larger heating rate
than a Jupiter-mass planet would, but the {\it possibility} of
inflation to that size is likely to be insensitive to where in the
planet heat is input (Bodenheimer et al. 2001), and thus insensitive
to the exact cause of the inflation.

\subsection{Results of the Simulations}

\begin{figure}
\plotone{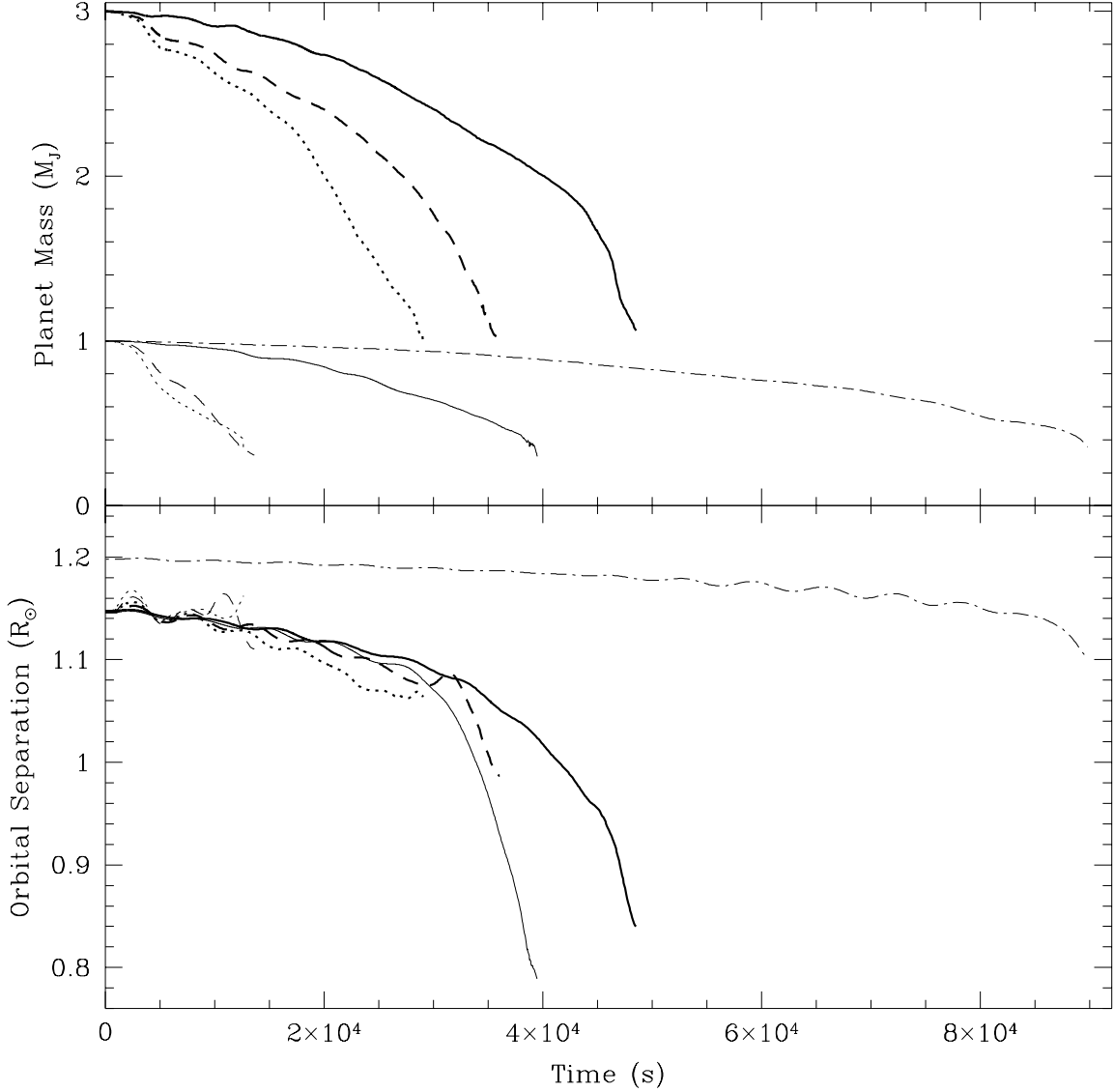}
\caption{(a) The temporal variation of planet mass for the
hydrodynamics simulations presented (in all cases, normal width lines
depict $1 M_{J}$ planets, and darker lines $3 M_{J}$ planets):
``cold'' ($R_{p} = 1 R_{J}$) planet ({\it solid line}); ``hot''
($R_{p} = 1.4 R_{J}$) planet and $1.2 M_{\odot}$ star ({\it dotted
line}); ``hot'' planet and $1.14 M_{\odot}$ star ({\it short-dashed
line}); ``hot'' planet, $1.2 M_{\odot}$ star, and higher start ({\it
long-dashed line}); ``cold'' planet, $1.2 M_{\odot}$ star, and higher
start ({\it dot-dashed line}).  The lines are terminated at the base of the
convection zone, or when the program is no longer able to track the
planet. (b) The temporal variation of the orbital
separation. \label{hydro}}
\end{figure}

The primary results are shown in Figure~\ref{hydro}.  Our simulations
of cold planets show, as expected, that they penetrate deeply into the
star before finally being destroyed. In our star models, the radius of
the convection zone base is $R_{cz} = 0.88 R_{\odot}$ so that a
Jupiter-mass planet was able to penetrate to the base of the star's
convection zone with approximately 50\% of its original mass.  So, if
much of the heavy element content (including Li) of the planet is in
its core, it would not be observable at the star's surface after the
planet's destruction. The cold $3 M_{J}$ planet also penetrates deeply
into the star, although the majority of the planet's mass is stripped
away before we were forced to end the run (when the code had
difficulty following the motion of the planet). Again, it is likely
that some portion of the planet's core would make it through the base
of the star's convection zone. By contrast, the hot giant planets were
much more easily ablated near the surface of the star. In all four
simulations (1 and 3 $M_{J}$ planets interacting with our 1.14 and
$1.2 M_{\odot}$ model stars), the planet is entirely destroyed before
penetrating through even half of the extent of the star's convection
zone.

The initial orbital decay timescales appear to be similar for all
runs, indicating that the timescale depends primarily on the
structure of the host star, which is similar for both of our stellar
structure models. A comparison with the cold planet simulations of
Sandquist et al. (1998) shows that the lower density of the stellar
envelopes in the current simulations results in longer timescales for
orbital decay.  Our results also show clearly that the orbits of the
cold planets take longer to decay on average than the hot ones because
the hydrodynamic drag is reduced for the more compact planets. The
majority of the orbital evolution occurs after the planet has already
lost a significant portion ($\sim 20 - 40$ \%) of its envelope mass.
Presumably, orbital evolution in the late stages of the interaction is
related to the amount of mass that is ablated and the amount of
momentum transferred.  The larger binding energy of the cold planets
results in delayed mass loss and slower orbital evolution while the
planet is still near the surface.

The orbital motions of the planets are also characterized by different
degrees of ``skipping'' across the star's surface --- an oblique shock
is quickly set up on the lower forward facing side of the planet. The
stellar gas flowing past the planet is thus able to transfer a
fraction of its momentum to the planet in the radially outward
direction, delaying orbital decay.

We conducted one additional simulation involving a cold Jupiter-mass
planet and a $M_\ast =1.2 M_{\odot}$ star, but with the planet
starting with only one-quarter of its diameter initially immersed in
the star surface. The lower initial drag forces resulted in a longer
timescale for orbital decay and a substantially longer
computation. The run hints that a cold planet hitting a star at a
shallow angle could be destroyed before reaching the base of the
star's convection zone. More realistic simulations would certainly
answer this question, but most factors (mass loss from the planet
prior to the hydrodynamic interaction, tidal distortion, etc.) seem to
favor planet destruction. Although simulation of the destruction of a
planet on an elliptical orbit is currently impractical due to long
interaction timescales, our results support the idea that heated
planets of small-to-moderate mass (and probably also cold planets)
could be destroyed relatively easily just below the star's
surface. Thus, the consumption of a heated planet is clearly a
plausible mechanism for polluting stars.

\section{Dynamical Origin and Evolution of a Doomed Short-Period Planet}

The basic assumption of the planet-consumption scenario is that some
metal-rich host SWPs may once have had close planetary companions
which migrated into their envelopes.  In order to enhance the
metallicity by a detectable amount, planet consumption must occur
after these stars have reached the end of their Hayashi track
pre-main-sequence evolution.  In addition, the dynamical stability of
the planetary system has to be considered. If we currently observe
some planets on stable orbits around a star that is suspected to have
been polluted, we have to monitor the effects of the time-varying
gravitational potential of the doomed planet before its destruction,
as well as the effects of the removal of its gravitational influence
afterward (for example, orbital resonances).  In this section, we
examine some processes which may be able to bring planets to the
surface of metal-rich SWPs.

While our objective is to provide a general assessment for the
planet-consumption scenario, our analysis is applied to the HD 82943
system because the dynamical constraint for this system is
particularly stringent.  The coexistence of two planets in a 2:1
resonance with periods of 221 and 446 d and eccentricities of $e=0.54$
and 0.41 have been reported (see
http://obswww.unige.ch/$\sim$udry/planet/hd82943syst.html).  The
requirement that this delicate configuration be preserved can be
used to constrain the initial conditions for the system.  The analytical
methods developed for this case can then be applied to other systems.

We first considered the possibility that the doomed planet was
gravitationally scattered off another giant planet onto an orbit that
brought it near the stellar surface as a consequence.  Multiple
planets are found around more than half of all the SWPs (Fischer et
al. 2001), and they form in a variety of configurations.  During the
epoch of their formation, the presence of residual gas in their
nascent disks may provide sufficient dissipation to preserve a stable
configuration.  But after the gas is largely removed, long-term
dynamical instabilities may lead to the excitation of large
eccentricities and induce planets to cross orbits in some systems
(Rasio \& Ford, 1996, Wiedenschilling \& Marzari 1996, Lin \& Ida
1997, Levison et al. 1998).  Subsequently, strong interactions and
close encounters between planets may scatter some of them close to the
surface or directly into the envelope of their host stars.  Although
this is a promising mechanism to account for the enhanced metallicity
of some SWPs, this scenario cannot be applied to HD 82943 because the
stability of the observed current 2:1 resonant system is finely tuned
in the sense that small changes in the orbital elements would render
this system unstable.  Therefore, we consider less disruptive avenues
of planetary migration.

\subsection{Tidal Dissipation within Doomed Planets}

During formation, tidal interactions between protoplanets and their
viscously-evolving nascent disk induce orbital
migration of the planets (Goldreich \& Tremaine 1980, Lin \& Papaloizou 1986).
Planets formed in the inner regions of the disk migrate toward their
host stars.  The migration of a given planet is halted if it
enters a magnetospheric cavity in the disk or if it undergoes tidal
interaction with a young, rapidly-rotating host star (Lin et
al. 1996).  These different mechanisms can both excite and damp the
incoming planet's eccentricity.  At such close range, the dissipation
of tidal energy within the planet's envelope leads to
strong damping of the planet's orbital eccentricity ($e$) and 
evolution of its spin frequency ($\Omega_p$) toward synchronization
with the orbital mean motion $n =(GM_\ast/a^3) ^{1/2}$.  This process
also heats the planet's interior and causes it to inflate (Bodenheimer
et al. 2000), and under some circumstances, to overflow its Roche lobe
(Gu et al. 2002).  But due to its small moment of inertia, very
little angular momentum is exchanged between the planet's spin and
its orbit.  In contrast, tidal dissipation 
within the envelope of a slowly/fast spinning host star
does lead to significant angular momentum transfer from/to the
planet's orbit to/from the stellar spin, and thus causes a
reduction/increase in the planets' semi-major axis $a$. In this
subsection, we assess the likelihood that these processes leading to
the disruption of the planet and contamination of the host star's
outer envelope.

First we examine the dominant effects of the tidal
dissipation in the planet's envelope, which induces
$\Omega_p$ to evolve towards $n$, and $e$ to damp on timescales
\begin{equation}
\tau_\Omega = {n \over \dot \Omega_p} = {\alpha_p Q_p^\prime \over n}
\left({M_\ast \over M_p} \right) \left({a \over R_p} \right)^3
\end{equation}
and
\begin{equation}
\tau_e = {e \over \dot e} = {4 \over 63} \left({M_p \over M_\ast} \right) 
\left({ a \over R_p} \right)^5 {Q_p^\prime \over n} 
\end{equation}
respectively,
where $R_p$ is the planet's radius, 
$\alpha_p = I / (M R^{2})$ its moment of inertia coefficient,
and $Q_p^\prime$ 
its effective dissipation parameter (Goldreich \& Soter 1966).  Since tidal dissipation
within the planet does not significantly modify its orbital angular
momentum $h=(G M_\ast a (1-e^2))^{1/2}$, the damping of $e$ leads to
an evolution in $a$ at a rate $\dot a = 2 a e {\dot e} / (1 - e^2)$.
In the absence of ongoing eccentricity excitation, the maximum change
in the semi-major axis is $\Delta a = - a e^2$.  Unless the planet is
on a nearly parabolic orbit, $\Delta a$ is not adequate to bring
it to the stellar surface from large distances.

\subsubsection{Planetary inflation and tidal disruption}
Eccentricity damping also leads to the tidal dissipation of the
planet's orbital energy $E = -G M_\ast M_p / 2 a$ which is deposited
as heat into the interior at a rate
\begin{equation}
\dot E_p = - {G M_\ast M_p \dot a \over 2 a^2} = {G M_\ast M_p e^2 
\over a (1-e^2) \tau_e}.
\end{equation}
In response, the gravitational binding energy of the planet is
reduced and the radius $R_p$ expands at a rate
\begin{equation}
\dot R_p = {2 \dot E_p R_p^2 \over G M_p^2 } = {2 M_\ast R_p^2 e^2 
\over M_p a (1 - e^2) \tau_e}.
\end{equation}
Provided 
\begin{equation}
a <2 (M_\ast/M_p) e^2 R_p/(1-e^2),
\label{eq:crita}
\end{equation} 
$\tau_R \equiv R_p/\dot R_p < \tau_e$ and the planet's envelope
expands before $e$ is damped out. A radiative equilibrium is
established for the planet at a critical size $R_c$ such that the planets' surface
luminosity ${\cal L}(R_c)$ is balanced by $\dot E_p$.  If $R_c$ is
larger than the Roche radius $R_R = (M_p/3M_\ast)^{1/3} a$, the
inflated planet begins to lose mass once $R_p \sim R_R$ (further
discussion of this process will be presented by Gu et al. 2002).  At
least a fraction of the lost mass flows through the L1 point and is
accreted onto the host star.

In the context of migration due to planet-disk interaction, the
necessary condition for planetary inflation in eq(\ref{eq:crita}) is
marginally attained as the migrating planet approaches its host star.
Provided the migration is halted at a modest value of $a$, the
planetary inflation timescale $\tau_R$ can be longer than the
timescale for the solar-type host stars to evolve onto the main
sequence ($\tau_c \sim 3 \times 10^7$ yr).  Since $M_p \ll M_\ast$,
the planet's tidal debris cannot significantly modify the host star's
metallicity until the mass contained in the star's convection zone is
reduced to a small fraction of $M_\ast$, which occurs on the timescale
of $\tau_c$ (Ford et al. 1999). In addition, the preservation of
\lisix is only possible after the star has reached the end of the
Hayashi track when the stellar convection zone has decreased to
approximately the present size of the solar convection zone (see
discussion in \S2).

For short-period planets that are scattered into the vicinity of the
host star as a consequence of dynamical instability in multiple-planet
systems, the condition for planetary inflation is not generally
satisfied immediately after the first close approach to the stellar
surface.  However, as $a$ and $e$ decline during the subsequent
circularization of the orbit, the condition in eq(\ref{eq:crita}) may
be satisfied.  Since this process is more likely to occur well after
the host star has evolved onto the main sequence, the stellar
convection zone can be noticeably polluted by tidally-disrupted stray
planets. Although this mechanism is a viable process for SWP
contamination, it is not likely to apply to the case of HD 82945
because the delicate two-planet orbital resonance would not have
survived, as we have indicated above.

\subsubsection{Eccentricity Excitation}

The orbital eccentricity of mature planets can also be excited long
after the nascent disk is depleted.  According to the pollution
hypothesis, metal-rich SWPs must have possessed two or more planets
originally.  Even today, multiple planet systems are indicated for
more than half of the SWPs (Fischer et al. 2001).  In some multiple
systems (such as $\upsilon$ And; Butler et al. 1999), short-period
planets coexist with intermediate- and long-period planets.  In such a
configuration, a short-period planet interacts both tidally with its
host star and dynamically with other planets. While tidal dissipation
within the planet removes energy from its orbit (Goldreich \& Soter
1966), secular interactions among planets transfer angular momentum
and excite eccentricity (cf. Murray \& Dermott 2000).  The combined
influence of these two effects is to cause the semi-major axis of the
innermost planet to decay on a timescale $\tau_{ea} \sim e^{-2}
\tau_e$.  (This process is outlined and analyzed in more detail in
Mardling \& Lin 2002b for the three planet system $\upsilon$ And,
where it appears to be important.)  Significant induced orbital decay
for the innermost members of multiple-planet systems would only be
possible if their periods are {\it long} enough for $e$ to be excited
to modest values and {\it short} enough for $\tau_e$ to be
considerably less than the age of the system.

Around mature solar-type main sequence stars (with ages \hbox{$\tau_\ast
\sim$ a few Gyr}), all Jovian-mass planets with periods less than 7 days
have nearly circular orbits.  Presumably, $\tau_e <$ a few Gyr for
these systems.  But for sensible choices of the planetary tidal
dissipation factor ($Q_p^\prime \sim 10^{5} - 10^{6}$; Yoder \& Peale
1981, Hubbard 1984), $\tau_{ea}$ of typical known short-period
Jupiter-mass planets would be longer than a few Gyr unless they can
maintain a modest or large orbital eccentricity.  In multiple-planet
systems, the magnitude of eccentricity variations due to secular
interaction is determined by the total energy and angular momentum,
the individual planet masses, and the precession frequency $\omega_s$,
which is a decreasing function of the ratio of the semi-major
axes. For short-period planets, the relativistic and rotational
corrections to the host star's gravitational potential, as well as the
potential due to the planet's tidal bulge, also cause
precession with a frequency $\omega_r$ (see Mardling \& Lin 2002b for a
comparison of these effects for $\upsilon$ And).
If they are accompanied by
other planets with comparable periods, $\omega_s > \omega_r$ and the
eccentricity is continually excited, driving orbital decay on a
timescale of $\tau_{ea}$.

For HD 82943, the observed planets have relatively long periods.  In
this type of system, $\omega_s < \omega_r$ and the amplitude of
eccentricity excitation is limited.  Nevertheless, during the epoch of
planet formation, the gravitational potential of the nascent disk
induces precession with a frequency $\omega_d$ that may be greater
than both $\omega_s$ and $\omega_r$ initially.  As $\omega_d$ declines
due to disk depletion, the eccentricities of several planets may be
excited as they evolve through a series of sweeping secular resonances
(Ward 1981 and Nagasawa \& Ida 2000 attribute the large eccentricity
distribution among asteroids and Kuiper Belt Objects to this process
of sweeping secular resonances).  Since the disk depletion timescale
$\tau_d (\sim$ a few Myr) is generally much shorter than $\tau_e$,
this effect cannot sustain an equilibrium eccentricity and
consistently removes angular momentum from the short-period planets.
Nevertheless, it can excite an initial eccentricity that may be large
enough to satisfy the conditions for planetary inflation and
Roche-lobe overflow.  As we have indicated above, the excited $e$'s
cannot be allowed to be too large and $a$'s too small or else the
system would have $\tau_R < \tau_c$ and the heavy element pollution,
including \lisix, would be deposited in the star while its convection
zone is still massive and able to effectively dilute the pollutants to
the point that they would be unobservable.

\subsection{Tidal Dissipation within the Host Stars} 

Spin periods measured from stellar variability or calculated from
\ion{Ca}{2} emission strength (Barnes 2001) indicate that most SWPs
have ``normal'' rotation frequencies $\Omega_\ast$ that are slower
than the orbital motions of short-period planets.  For slowly spinning
stars, the dissipation of the tidal disturbance raised by a
short-period planet tends to spin up the star at the expense of the
planet's orbital angular momentum such that $a$ decreases on a timescale
\begin{equation}
\tau_a = {a \over \vert \dot a \vert } = {Q_\ast^\prime \over 2 n}
\left( {M_\ast \over M_p} \right) \left( {a \over R_\ast } \right)^5
\end{equation}
where $R_\ast$ is the radius of the host star (Goldreich \& Soter
1966). The magnitude of the host star's \hbox{$Q_\ast^\prime$-value}
can be determined from models of either equilibrium or dynamical tides
(cf.  Zahn 1989, Goldreich \& Nicholson 1989, Terquem et al. 1998,
Goodman \& Oh 1999). With these theoretical $Q_\ast^\prime$ values,
the inferred $\tau_a$ for Jupiter-mass planets around solar-type stars
is generally longer than several Gyr unless $P< 8-9$ h (Rasio et
al. 1996, Marcy et al. 1997).  However, these theoretical $Q^\prime$
values also imply a slow rate of circularization for the orbits of
close solar-type binary stars, which is inconsistent with observations
of binaries in stellar clusters of various ages (Mathieu 1994).  Using
the observed circularization timescales as a calibration, the
inferred $Q_\ast^\prime \sim 1.5 \times 10^5$ for young stellar
objects (Lin et al.  1996) and $\sim 2 \times 10^6$ for mature stars
(Terquem et al. 1998).  For these values of $Q_\ast ^\prime$, $\tau_a$
of planets with $a <0.025$ AU is comparable to or less than the main
sequence lifespan of their solar-type host stars.  (This value depends
slightly on the mass of the hypothetical planet, with an upper limit
imposed by stability considerations in some multiple-planet systems;
see next section). Thus this is a viable mechanism for the
consumption of short-period planets by their host stars.

Although we have made allowance for angular momentum transfer between
planets and their host stars, we have so far neglected any changes in
the total angular momentum of the star-planet system.  We have already
addressed the issue of angular momentum transfer between the planets,
their host stars, and their nascent disks.  On a timescale comparable
with $\tau_c$ (a few Myr), the protostellar disks are depleted,
resulting in a decline in the rates of angular momentum drainage from
the systems to the disks.  However, solar-type main sequence stars
also lose angular momentum as their age $t_\ast$ increases such that
their surface rotation frequency $ {\Omega_\ast / \Omega_o} \simeq
({t_\ast / t_o})^{-1/2}$, where $\Omega_o = 10\, {\rm km} \ {\rm
s}^{-1} / R_\ast$ and $t_o = 2 \times 10^8$ yr (Skumanich 1972,
Soderblom 1983).  The above scaling law implies a spin-down rate $\dot
\Omega_{-} \simeq (\Omega_o/2 t_o) (\Omega_\ast / \Omega_o)^3$.  Tidal
interaction also leads to angular momentum exchange between the
star's convective envelope and the planet that modifies $\Omega_\ast$
at a rate
\begin{equation}
\dot \Omega_t \simeq {\rm sign} (n-\Omega_\ast) {M_p a^2 \over M_{cz}
R_\ast^2 } {n \over 2} {\dot a \over a}.
\end{equation}
The total rate of angular frequency evolution is $\dot \Omega_\ast =
\dot \Omega_t - \dot \Omega_{-}$.  

$[$In arguments throughout this paper, we assume that the star's
convection zone and its radiative interior rotate independently of
each other. The most direct evidence of this comes from
helioseismology through frequency splitting of the Sun's oscillation
modes (Thompson et al. 1996), showing the differentially rotating
convection zone surrounding the uniformly and more rapidly rotating
interior. Observations of young open cluster stars (Soderblom et
al. 1993) suggest that the radiative interior of solar-type stars
provides a long-term source of angular momentum for the
convection zone, maintaining the surface rotation at modest values ($v
\sin i \sim 10$ km s$^{-1}$). Based on these considerations, we believe
the assumption of a rotationally-decoupled convection zone is
realistic.$]$

Shortly after disk depletion, a short-period planet migrates outward
slightly as its host star spins down.  For $a > a_c \equiv
(M_{cz}/M_p)^{1/2} R_\ast$, the moment of inertia of the planet's orbit is
larger than that of the star's convective envelope so that when the
star reaches synchronization with the planets' orbit it is secularly
stable to conservative tidal transfer of angular momentum
(Chandrasekhar 1968, Hut 1980, Lin 1981, Rasio et al.\ 1996).  But as
the host star continues to spin down the planet can resume its inward
migration (Bryden et al. 1998).  A quasi-equilibrium value $\Omega_e
\sim (M_p t_o n /M_{cz} \tau_a \Omega_o)^{1/3} (a/R_\ast)^{2/3}
\Omega_o$ is attained when $\dot \Omega_t \sim \dot \Omega_{-}$.  The
initial value of $\Omega_e R_\ast \sim$ a few km s$^{-1}$ when $\tau_a
\sim 1$ Gyr because of the strong dependence of the spin-down rate on
surface rotation frequency. The stars are thus able to rid themselves
of most of the angular momentum transferred to them by their planets
until their spin rate gets low enough.  Based on this argument, we can
safely assume a small initial value for $\Omega_\ast (= 2 \pi / $ 500
h) in our analysis.

The magnitude of $\dot \Omega_t$ increases rapidly as $a$ decreases.
Close to the star, $\dot \Omega_t > \vert \dot \Omega_{-} \vert$ so
that $\Omega_\ast$ would increase with time.  But $\Omega_\ast$ would
be less than $n$ and the planet's orbit would continue to decay if $a
< a_c = (M_{cz}/M_p)^{1/2} R_\ast$.  Even for $a > a_c$, the loss of
angular momentum from the system eventually leads $\Omega_\ast$ to be
much smaller than that expected from conservative tidal evolution
$\Omega_c = (M_p / (M_p + M_{cz})) (a_i / R_\ast) ^{1/2} \Omega_s$,
where $a_i$ is the initial value of $a$ and $\Omega_s =
(GM_\ast/R_\ast^3) ^{1/2}$ is the Keplerian frequency near the stellar
surface.  For numerical models 1 and 2 which will be presented below,
we choose, respectively, $\Omega_c / \Omega_s = 0.1$ and $0.5$ for the
innermost planet --- in other words, the planet's orbital decay is
unlikely to be impeded by the attempted synchronization of the spin of
the stellar surface with the planet's mean motion. Note that the spin
of the stellar convection zone is unlikely to modify the gravitational
potential and introduce additional planetary precession because of the
small values of both $\Omega_c / \Omega_s$ and $M_{cz}/M_\ast$.  Thus,
the relative speed of the orbiting planet and the spinning convective
envelope is expected to be large, leading to the initial state of our
dynamical models and hydrodynamical simulations.

\subsection{Dynamical Stability of a Three-Planet System}

If the contamination scenario is to be able to explain the high metal
abundances in a large fraction of SWPs, it will be necessary to
require some of the planetary systems to have stable (at the very
least, dynamically) configurations prior to and after the planet
consumption event. For those SWPs that contain one or more
intermediate- or long-period planets today, the surviving planets can
have been dynamically decoupled from the hypothesized short-period
planet such that its existence and disruption had little dynamical
consequence. For systems containing short-period planets
today, even though the surviving planets may have intensely interacted
with their demised siblings, the signatures of the encounters can be
erased by their own subsequent tidal interactions with the host star.
However, systems containing resonant planets, such as HD 82943, may be more
sensitive to perturbations by the hypothesized short-period planet.
Therefore it is important to assess the requirements for
coexistence.  In this subsection, we carry out numerical simulations
to establish the existence of stable and perhaps dynamically
decoupled systems, which contain both short-period and resonant
planets, and then follow their dynamical evolution, taking into
account the effects of tidal and secular interactions as well as
stellar spindown.

In order to quantitatively determine the relative importance of these
processes, we used the (direct) numerical model described in Mardling
\& Lin (2002a) for all our dynamical studies. This code can be used to
study the dynamical evolution of up to three planets, including the
secular effects of the relativistic potential of the star and the
tidal evolution and spin-orbit coupling of the star and innermost
planet. Orbit averaging was not used because, among other things, we
were interested in stability issues.

A comprehensive stability analysis requires an extensive simulation
for a large number of orbital parameters such as the planets' masses,
semi-major axis, and eccentricity distribution.  Such a study is
beyond the scope of the present paper.  Here we use the specific
example of HD 82943 to demonstrate that resonant planets can easily
retain their observed dynamical properties despite the initial
existence and subsequent disruption of hypothetical short-period
planets.

An understanding of the stability of the observed resonant system
including the resonant capture process deserves its own dedicated
study (for a study of the other known, and much shorter period, 2:1
resonant system GJ 876, see Lee \& Peale 2001).  The high
eccentricities and similar masses of the two planets orbiting HD 82943
present a challenge to existing theories of stability and resonant
capture (Murray \& Dermott 2000).  From an empirical point of view, it
is possible to constrain the configuration parameters using
observational data as has been done for GJ 876 (Laughlin \& Chambers
2001). In lieu of such a study, we chose our working orbital
parameters as follows.  We started by considering the stability of the
observed two-planet system.  Using the orbital parameters available at
the time of writing ($e_1=0.54\pm0.05$, $e_2=0.41\pm0.08$, $a_1=0.73$
AU, $a_2=1.16$ AU, $\varpi_1$ not determined, $\varpi_2=117.8^o$,
where subscripts 1 and 2 refer to the inner and outer planets
respectively and $\varpi$ is the longitude of periastron), together
with the minimum masses ($M_1\sin i=0.88 M_J$, $M_2\sin i=1.63
M_J$), and zero internal inclination (in the absence of a measured
estimate), there exist stable configurations only for {\it some}
values of the initial relative phases, at least for the short times we
integrated the orbits ($10^4$ yr). However, these ``stable'' systems
exhibit large variations in the eccentricities of both planets, as
well as large variations in the relative longitude of periastron (for
whatever value we choose for the initial value of $\varpi_1$). Our
study of GJ 876 (Lin \& Mardling, in preparation) together with the
theory of resonant capture (Murray \& Dermott 2000) lead us to believe
there should be very little variation in these parameters, and hence
we expect future observations of the HD 82943 system to reveal
slightly different values for the eccentricities.

\begin{figure}
\plottwo{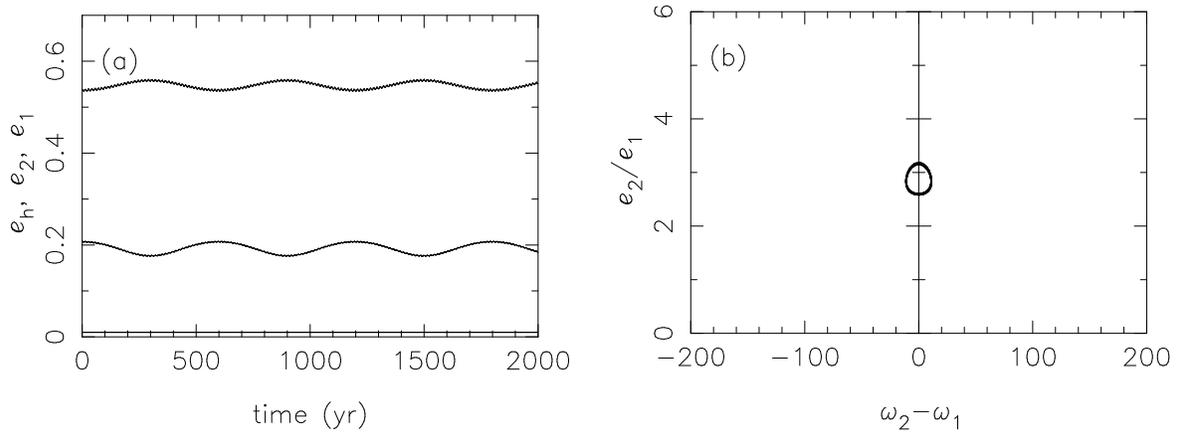}{f3b.eps}
\caption{Model 1: (a) Secular variation of the eccentricities.
The subscript $h$ 
refers to the hypothetical planet while 1 and 2 refer to the
inner and outer observed planets respectively.
(b) Variation of the relative longitude of 
periastron. Note the symbol $\omega$ is written as $\varpi$
in the text.
\label{fig3}}
\end{figure}

For the present study, we chose to vary the outer eccentricity in
order to produce a system with the above properties.  To do this we
used a formula derived by Mardling \& Lin (2002a; equation~9) which
gives the critical ratio of eccentricities for which there is no
secular variation in the eccentricities and relative longitude of
periastron for systems of two planets.  This formulation does not take
into account any variations due to resonant effects, but nonetheless
yields a system with essentially the desirable properties.
Figure~\ref{fig3} illustrates this for a three-planet system with
$e_2=0.207$. The addition of a short-period planet at 0.02 AU ($M\sin
i=0.5 M_J$) does not disturb the stability, nor does it affect the
extent of variations of the orbital parameters of the two outer
planets. In this configuration, provided $M_p < 2.5 M_J$, the secular
evolution of the outer two planets proceeds as if the innermost planet
did not exist; similarly the orbit of the innermost planet decays on
the same timescale it would were it isolated. The orbit of the
innermost planet is also unaffected by the secular perturbation of the
outer two resonant planets because the relativistic correction
dominates the precession of its periastron longitude.  Although we
cannot investigate the long-term evolution of this system with our
numerical code, we are confident that stability of the system persists
because the innermost planet becomes more and more decoupled from the
outer two as time goes on.  It is likely that Nature has arranged for
the observed system to suffer even less secular variation than
illustrated here; however our hypothetical system serves the purpose
of the present discussion.

In order to be consistent with our hydrodynamical simulations, we
considered a range of masses for the hypothetical short-period planet,
with the eccentricity taken to be 0.01 in each case and the spin
period of the star taken to be 500 h (the solar value).  For the
present calculation we adopt $Q_\ast^\prime = 10^6$ so that given a
planet with $M_p=0.5 M_J$, the initial semi-major axis needed to
achieve the prescribed decay timescale ($\tau_a \sim$ 1 Gyr) is
$a_i=0.02$ AU (model 1; see Figure~\ref{fig3}), while a planet of $2.5
M_J$ can start a little further out at $a_i=0.025$ AU (model 2).
These simulations show that 1) the orbit of the innermost planet does
indeed decay on the timescale $\tau_a \sim 1$ Gyr, 2) the resonant
configuration of the outer two planets is preserved despite the demise
of the hypothetical planet, and 3) the host star can lose sufficient
angular momentum to account for its observed slow spin.

\section{Summary and Discussions}

The hydrodynamical simulations presented here indicate that if a
heated giant planet can be brought into contact with the stellar
surface, it is likely that the planet would be easily destroyed, and
its ablated material would be thoroughly mixed in the star's
convection zone, thereby modifying the surface abundances in an
observable way.  Such a planet may have formed several AU from its
host star and migrated inward as a consequence of tidal interaction
with the protoplanetary disk shortly after formation.  This early
phase of migration may be halted if the planet enters a magnetospheric
cavity near the inner boundary of the disk or if it interacts tidally
with a slowly rotating host star.  If it arrives near the stellar
surface with a modest eccentricity, such a planet may be disrupted
outside the star as a consequence of intense tidal dissipation on a
timescale longer than the time needed for the star to reach the main
sequence.  But if it arrives near the stellar surface with an almost
circular orbit, the planet may resume its inward migration as its host
star's spin frequency declines below its orbit frequency.  The driving
force in this latter scenario is the dissipation of the planet's tidal
disturbance in the star. This tidal interaction can cause the planet
to eventually enter into the atmosphere and envelope of its host star
while the star is still on the main sequence.  A substantial portion
of a cold planet's mass (most importantly, a core) may survive the
plunge, although a definitive answer will require more sophisticated
simulations. Regardless, the destruction of a cold planet in a star
could still pollute the star if the planet has little or no rock/ice
core (in other words, if its heavy elements are mixed throughout the
planet).

If a giant planet mostly disintegrates within the convection zone of
its host star, the momentum input from the planet would increase the
rotational velocity of the star's convection zone above $M_p \Omega_s
R_\ast / (M_p + M_{cz})$ if the convection zone rotates as a solid
body.  At the interface between the convective and radiative zones,
there is likely to be a large shear as a result. This shear could in
principle result in mixing between the convection zone and interior,
and thus would tend to weaken any pollution signature. To gain some
guidance as to whether this could occur, we can look at the Sun
(Thompson et al. 1996). For the Sun, a tachocline transition layer
(Spiegel \& Zahn 1992) separates the surface convection zone and the
radiative interior, which have distinct rotational properties.  It has
been suggested that an interior magnetic field is responsible for
confining the shear in the tachocline and quenching mixing between
these regions (Gough \& McIntyre 1998).  Extrapolating from these
solar phenomena, it it possible that the rotational and compositional
mixing between the convection zone and the radiative interior of SWPs
with masses near that of the Sun would be limited.  Application of the
stellar spin-down formula would imply that the surface rotational
speed may be reduced substantially in less than a Gyr, limiting the
duration of any mixing that might occur.

In stellar evolution, atomic species like \lisix, \lisev, \benine, and
B all provide valuable diagnostics of the evolution of surface
convection zones, so it is worthwhile to try to understand how much
information they can reveal about the frequency of pollution among
planetary systems. The temperatures at the base of the convection
zones of pre-main sequence stars result in complete depletion of
\lisix and moderate depletion of \lisev, while \benine is mostly
unaffected. \lisix abundances constrain the pollution that can have
occurred since an age of about 30 Myr after star formation.  We point
out that the amount of \lisix detected in a star could distinguish
between {\it different} pollution mechanisms if giant planets are
formed around massive planetesimal cores, and that this test is mostly
independent of giant planet mass.

The diagnostic usefulness of \lisix and the other species is limited
for low-mass stars ($M \lapprox 1.1 M_{\odot}$) because their
convection zones are too massive, diluting pollutants to the point
that they are unobservable. Unfortunately, these stars are also the
ones that have the most nuclear processing of diagnostic species, and
thus would be easiest to distinguish from an unpolluted star. Stars
with $1.1 M_{\odot} \lapprox M \lapprox 1.3 M_{\odot}$ have relatively
shallow convection zones that are easily polluted, but this also
reduces the temperature at the base of the convection zone to the
point that the depletion rates for the lithium isotopes are
small. There is some evidence for \lisev depletion (and therefore,
presumably \lisix) in this range from comparisons of stars in the
Pleiades, Hyades, and older clusters like NGC 752 and M67.  \benine
and B are unlikely to be affected.

Stars with masses a little above $1.3 M_{\odot}$ are not currently
understood.  Standard models predict a small surface convection zone
or none at all, while stars in the Hyades (a very well-studied
cluster) show severe \lisev depletion (the so-called ``Li
dip''). Comparison with observations of other open clusters indicates
that the depletion timescale for stars in the dip is less than about
the age of the Hyades ($\sim 600$ Myr). Though a theoretical
understanding of the depletion mechanism is currently lacking (see
Pinsonneault 1997), stars that would normally fall in the Li dip seem
to be excellent candidates for a search for evidence of
pollution. Deposited \lisev would be expected to survive for a large
enough fraction of a SWP's main sequence lifetime ($\sim 30\%)$ that
the chances of detecting a pollution signature are favorable. This
would be best accomplished using comparisons with open clusters of age
similar to the host star. Murray \& Chaboyer (2002) have recently
asserted that the [Fe/H] abundances for SWPs in the Li dip can be
explained by hypothesizing a mixed layer at the star's surface of the
size needed to explain the \lisev depletion pattern. If true, a
pollution signature would of course be diluted. In either case, the
diagnostic possibilities are good, and so we would like to add our
voices to encourage observations of SWPs in this range.

Finally, we present a scenario which may account for the dynamical
history of the system HD 82943.  We show that the resonant planets
around HD 82943 are dynamically stable if the inner and outer planets
have eccentricities $0.54$ and $0.207$ respectively.  In this
scenario, although the eccentricity of the inner resonant planet is
the same as observed, that of the outer resonant planet is
considerably smaller than that reported.  We find that a planet with a
mass $0.5-2 M_J$ and a semi-major axis $0.02-0.025$ AU is
dynamically decoupled from the resonant planets with semi-major axes
of 0.73 and 1.16 AU.  The orbital configuration of this planetary
system is similar to that of $\upsilon$ And, and it is formed
naturally as a consequence of planet-disk interaction.  Under the
action of tidal orbital evolution, this hypothetical short-period
planet may have undergone orbital decay into the host star on a timescale less than 1 Gyr.  The analogues of such a planetary system might
be observed around young stars after the T Tauri phase of the star's
evolution.  The existence of short-period planets around relatively
young stars with normal metal abundance is a natural prediction of
this scenario.  A direct search for such systems is technically
feasible with the Space Interferometer Mission.

\acknowledgments 

We thank A. Burkert for the use of his nested-grid hydrodynamic code.
We would like to thank B. Jones, P.-G. Gu, P. Bodenheimer, B. Fegley,
D.O.  Gough, and D. Lambert for helpful conversations.  This work is
supported in part by NASA through grant NAG5-10727 and NSF through
AST-9987417 to D.N.C.L., and J.J.D. was partly supported by an RSCA
grant from San Diego State University to E.L.S.

\clearpage
 
\clearpage
 
\clearpage
 
\clearpage

\begin{deluxetable}{ccl}
\tablewidth{0pt}
\scriptsize
\tablecaption{Observational Constraints for HD 82943}
\tablehead{
\colhead{Observation}&
\colhead{Value}&
\colhead{Reference}}
\startdata
$M_{V}$ & $4.36 \pm 0.05$ &  GCPD\tablenotemark{a}, Hipparcos\\
$T_{eff}$ & $6010 \pm 50$ K & Santos et al. 2000\\
 & $6025 \pm 40$ K & Santos et al. 2001\\
$[$Fe/H$]$ & $+0.32 \pm 0.06$ & Santos et al. 2000\\
 & $+0.33 \pm 0.06$ & Santos et al. 2001\\
$\log g$ & $4.62 \pm 0.20$ & Santos et al. 2000\\
 & $4.54 \pm 0.10$ & Santos et al. 2001\\
\multicolumn{3}{c}{Derived}\\
$\log (L / L_{\odot}) $ & $0.18\pm0.02$ & \tablenotemark{b} \\
\enddata
\label{obs}
\tablenotetext{a}{The General Catalogue of Photometric Data; http://obswww.unige.ch/gcpd/gcpd.html; Mermilliod, Mermilliod, \& Hauck 1997}
\tablenotetext{b}{Uses bolometric correction of $-0.07$. Error does not include
possible systematic error in bolometric correction.}
\end{deluxetable}

\begin{deluxetable}{lcc}
\tablewidth{0pt}
\scriptsize
\tablecaption{Stellar Models for HD 82943}
\tablehead{
\colhead{Quantity }&
\colhead{Unpolluted Model}&
\colhead{Polluted Model}}
\startdata
$M$ & $1.2 M_{\odot}$ & $1.14 M_{\odot}$\\
Age & 1.05 Gyr & 1.5 Gyr\\
$R$ & $1.148 R_{\odot}$ & $1.143 R_{\odot}$\\
$T_{eff}$ & 6015 K & 6025 K\\
$L$ & $1.526 L_{\odot}$ & $1.525 L_{\odot}$\\
$[$Fe/H$]_{f}$ \tablenotemark{a} & $+0.29$ & $+0.22$\\
$\log g$ & 4.397 & 4.397 \\
$R_{cz}$ \tablenotemark{b} & $0.7662 R$ & $0.7667 R$\\
$M_{cz}$ \tablenotemark{c} & $0.0109 M_{\odot}$ & $0.0082 M_{\odot}$\\
\enddata
\tablenotetext{a}{Metallicity at the observed age}
\tablenotetext{b}{Radius of the base of the surface convection zone}
\tablenotetext{c}{Mass in the surface convection zone}
\label{smods}
\end{deluxetable}

\end{document}